\documentclass{PoS}

\usepackage{url}
\usepackage{graphicx}
\usepackage{amsmath}
\usepackage{slashed}
\usepackage[utf8]{inputenc}

\def\beq{\begin{equation}}
\def\eeq{\end{equation}}
\def\bea{\begin{eqnarray}}
\def\eea{\end{eqnarray}}
\def\D0{D\O }

\title{Magnitudes of $V_{xb}$ CKM matrix elements}

\ShortTitle{$|V_{xb}|$ Determination}

\author{\speaker{Giulia Ricciardi}%
\\
        Dipartimento di Scienze Fisiche, Università   di Napoli Federico II\\
        E-mail: \email{giulia.ricciardi@na.infn.it}}

\abstract{We review the current status  of the absolute values   of the CKM matrix elements $V_{xb}$, with
 particular attention to latest determinations.
}

\FullConference{14th International Conference on B-Physics at Hadron Machines\\
		 April 8-12, 2013\\
		 Bologna, Italy}

\begin{document}

\section{Introduction}

The Cabibbo-Kobayashi-Maskawa (CKM) matrix
parameterizes the couplings
between flavors of quarks under  weak interactions. The values of the CKM matrix elements are not predicted by the Standard Model (SM), and, in the last decade, a large effort has  gone towards their determination, driven by increasingly higher statistics at new and improved facilities and by theoretical advances.
We summarize the current status  of the absolute values   of the CKM matrix elements $V_{xb}$, with
 particular attention to latest developments.
For  recent theoretical  reviews and for  notations, see e.g. \cite{Ricciardi:2013cda, Ricciardi:2012pf}.

\section{The $|V_{cb}|$ determination}

 Currently, $|V_{cb}|$ is  inferred from semi-leptonic decays where the final lepton is an electron or a muon, $b \to \tau$ semileptonic decays being mostly studied for their sensitivity to New Physics (NP).
In the massless limit, the exclusive $ \bar{B}\rightarrow D^{(\ast)}  l \bar{\nu}$ decays depend only on one form factor ${\cal G}  (\omega)$ (${\cal F}  (\omega)$), where $\omega$ is the product of the velocities   of the hadrons in the HQET framework. The  determination of non-perturbative contributions to the form factors is the major theoretical challenge.  The $ \bar B \to D^\ast l \bar \nu $ decay is more advantageous for the exclusive estimate of  $|V_{cb}|$,  since it occurs at an higher  rate than $\bar  B \to D l \bar \nu $, and  non-perturbative,  linear corrections to the form factor  ${\cal F}$ are absent at zero recoil.
Lattice corrections to  ${\cal F}  (\omega)$ at finite momentum transfer ($\omega= 1.075$) are available in the
quenched approximation  \cite{deDivitiis:2008df}; combined with 2008 BaBar data \cite{Aubert:2007rs} they give
\beq
|V_{cb}| =  (37.4 \pm 0.5_{\mathrm{exp}} \pm 0.8_{\mathrm{th}} ) \times 10^{-3}
\label{latticeTV}
\eeq
with a rather small nominal error.
Lattice corrections to  ${\cal G}  (\omega)$ at the same finite momentum transfer are also available, in the
quenched approximation \cite{deDivitiis:2007ui, deDivitiis:2007uk}. By
 using 2009  BaBar  data \cite{Aubert:2009ac}, a slightly higher value is found
 \beq |V_{cb}| = (41.6 \pm 1.8 \pm 1.4
\pm 0.7_{\mathrm{FF}} ) \times  10^{-3} \label{lattunq1} \eeq
The errors are  statistical, systematic and due to the theoretical uncertainty in the form factor $ {\cal G}$, respectively.
Unquenched calculations are only available at $\omega=1$, and  have been performed   by the FNAL/MILC Collaboration. The   Heavy Flavor Averaging
Group  (HFAG) experimental  averages \cite{Amhis:2012bh}, combined with the latest update for  ${\cal F}  (1)$ \cite{Bailey:2010gb},  give
\beq
|V_{cb}| =  (39.54 \pm 0.50_{\mathrm{exp}} \pm 0.74_{\mathrm{th}} ) \times 10^{-3}
\label{lattice53}
\eeq
and
 \beq |V_{cb}| = (39.70 \pm 1.42_{\mathrm{exp}} \pm 0.89_{\mathrm{th}})
\times  10^{-3} \label{lattice2} \eeq
 combined with  the less recent  unquenched lattice results for ${\cal G}  (1)$ \cite{Okamoto:2004xg, Laiho:2005ue},
The two values are
 in good agreement, although the theoretical error in the determination based on $ \bar B \to D l \bar \nu $ decays is slightly larger and
the experimental one more than twice larger.
Let us observe that  a further  update of  ${\cal F}  (1)$  by  FNAL/MILC Collaboration  has  been announced   \cite{JackLaihotalk}, claiming a reduction of discretization effects and of the error on $|V_{cb}|$ down to 1.6\%, but no new value for $|V_{cb}|$ has been published until now.
Also  studies  of the form factor  ${\cal G}  (\omega)$   at non-zero recoil are in progress, some preliminary results being already available \cite{Qiu:2012xi}.
On the non-lattice side,
recent estimates for ${\cal F}  (1)$  have  been  reported in the zero-recoil sum rules framework \cite{Gambino:2012rd},  yielding  to
\beq
|V_{cb}| =  (41.6\pm 0.6_{\mathrm{exp}}\pm 1.9_{\mathrm{th}}) \times 10^{-3}
\label{VCBF1}
\eeq
combined with the HFAG experimental data fit \cite{Amhis:2012bh}.
The theoretical error is more than twice the error of the lattice determination given in Eq. (\ref{lattice53}), but
the budget error from lattice has been  questioned~\cite{Gambino:2012rd}. The claim is that existing  differences between the power-suppressed deviations from the heavy flavour  symmetry
 in
the lattice theory with heavy
quarks and in continuum QCD may be compensated by a matching between the two theories that has  been performed, at the best, only at
lower levels.
The most recent non lattice calculation of non-perturbative contributions to ${\cal G}  (1)$ dates 2004 and  combines the heavy quark expansion
with expanding around the point
where the kinetic energy
 is equal to
the chromomagnetic moment
 $\mu_\pi^2=\mu^2_G$ (''BPS" limit)~\cite{Uraltsev:2003ye}.
With this estimate, the PDG finds \cite{Beringer:1900zz}
\beq
|V_{cb}| =  (40.7 \pm 1.5_{\mathrm{exp}} \pm 0.8_{\mathrm{th}}) \times 10^{-3}
\eeq
in agreement, within the errors, with both lattice
determinations  (\ref{lattunq1})   and  (\ref{lattice2}).
Semileptonic $B$ decays to orbitally-excited P-wave
charm mesons ($ D^{\ast \ast}$)
contribute as a source of systematic error in the $|V_{cb}|$ measurements at the B factories (as
previously at LEP), as a background to the direct decay $ B^0 \to D^{\ast } l  \nu$.
The  knowledge of  these semileptonic
decays is not complete yet: one example for all,  the so called "1/2 versus 3/2" puzzle.
Very recently,  dynamical lattice computations of the $ B \to D^{\ast \ast } l \nu$ form factors have been attempted, although  still preliminary and needing extrapolation to the continuum~\cite{Atoui:2013sca}.

In most of the phase space for  inclusive $ B \rightarrow X_q  l  \nu$ decays,  long and short distance dynamics are factorized by means of  the heavy
quark expansion.
However, the  phase space region includes a  region of singularity, also called endpoint or threshold region, plagued by the presence
 of large double (Sudakov-like)  perturbative  logarithms at all orders in the strong coupling\footnote{for
 theoretical aspects of threshold
resummation in $B$ decays
see e.g.~\cite{Aglietti:2002ew, Aglietti:2005mb, Aglietti:2005bm, Aglietti:2005eq, Aglietti:2007bp, DiGiustino:2011jn}.} and by  enhanced non-perturbative effects.
The  $b \rightarrow c$ semileptonic decays are not affected by the small region of singularity in a significant way; in addition,  corrections are not expected  as singular as in the $ b \rightarrow u$ case, being  cut-off by the charm mass.
Recently,  a global fit~\cite{Amhis:2012bh} has been  performed to the width and all
available measurements of moments in $ B \rightarrow X_c  l  \nu$ decays, yielding, in the kinetic scheme
\beq |V_{cb}| = (41.88 \pm 0.73) \times 10^{-3} \eeq
and in the 1S scheme   \beq |V_{cb}| = (41.96 \pm 0.45) \times 10^{-3} \eeq
Each scheme has its own non-perturbative parameters that have been estimated together with the charm and
bottom masses. The inclusive
averages are in substantial agreement with the values extracted from exclusive decays, within the errors.

\section{The $|V_{ub}|$ determination}

The analysis of
exclusive charmless semileptonic decays is currently employed to determine the CKM parameter $|V_{ub}|$, which
 plays a crucial role in the study  of
the unitarity constraints.
Also here,  information about hadronic matrix elements is required via form factors.
Among all possible channels,
the  $ B \rightarrow \pi l  \nu$ mode benefits of more precise branching fraction measurements  and is currently the  channel of election
 to
determine  $|V_{ub}|$ exclusively. In the limit of zero leptonic masses, it is
 affected by a single form factor $f_+(q^2)$,
The  first lattice determinations of  $f_+(q^2)$  based on unquenched  simulations have been obtained by the FNAL/MILC \cite{Bailey:2008wp}  and the HPQCD \cite{Dalgic:2006dt} collaborations and  are  in substantial agreement.
Recent results are also available on a fine lattice (lattice spacing $a \sim 0.04$ fm) in the quenched approximations by the QCDSF collaboration \cite{AlHaydari:2009zr}.
The measured
partial $B \rightarrow \pi l  \nu$ branching fractions can be directly fit at low and high $q^2$  according to light-cone sum rules (LCSR) and lattice approaches, respectively, the latter providing generally better fits.
Alternatively, theoretical extrapolations to the full $q^2$ range can be employed, in a simultaneous
fit to theoretical  results  and
experimental data.
The most recent simultaneous fit to the data over the full $q^2$ range and the FNAL/MILC
 lattice  results  \cite{Bailey:2008wp} has been performed by  BaBar \cite{Lees:2012vv} and has
given  the following average value
\beq
|V_{ub}| = (3.25 \pm 0.31) \times 10^{-3} \label{exclus}\eeq
in agreement with the  analogous fit by Belle \cite{Ha:2010rf}.
The $|V_{ub}|$ determination inferred
by  using lattice QCD
 direct calculations,
 in the kinematic region $q^2 > 16$ GeV$^2$, with no extrapolations,
is also in agreement with the previous value.
In the complementary   kinematic region, at
large recoil,  with an upper limit for $q^2$ varying between 6 and 16 GeV$^2$,
 direct LCSR calculations of the semileptonic  form factors  are available, which have benefited by  recent progress in pion distribution amplitudes, NLO and LO higher order twists (see e.g.~\cite{Khodjamirian:2011ub, Bharucha:2012wy,Li:2012gr} and Refs. within). The $|V_{ub}|$ estimate are generally higher than the corresponding lattice ones, but still in agreement, within the relatively larger theoretical errors.
The latest determination, from BaBar Collaboration \cite{Lees:2012vv},
 using  LCSR  results   below $q^2 = 16 $ GeV \cite{Khodjamirian:2011ub}, gives
\beq |V_{ub}| =
(3.46 \pm 0.06 \pm 0.08^{+0.37}_
{-0.32})  \times 10^{-3}  \eeq
 where the  three uncertainties  are statistical, systematic and theoretical, respectively.
A consistent value has been found by the  Belle collaboration \cite{Ha:2010rf}, using a different LCSR   form factor determination  \cite{Ball:2004ye}.

Recently, BaBar and Belle Collaborations have  significantly improved the branching ratios of other
 heavy-to-light semileptonic decays,  yielding to an
increased precision
for $|V_{ub}|$ values inferred by these decays \cite{Lees:2012vv}. The $|V_{ub}|$ values extracted
from  $ B^+ \rightarrow \omega l^+ \nu$ \cite{Lees:2012vv} and $ B \to \rho l \nu$ \cite{delAmoSanchez:2010af} agree  with $|V_{ub}|$  determinations   inferred from the $ B \to \pi l \nu$ decay mode within the errors, which  for the $ B \to \rho l \nu$  channel are
starting to become comparable.
Other interesting  channels  are
 $ B \rightarrow \eta^{(\prime)} l \nu $ \cite{delAmoSanchez:2010zd, DiDonato:2011kr}, but  a value of
$|V_{ub}|$ has not been extracted  because
the theoretical partial decay rate is not sufficiently
precise yet.
There has been  also  recent progress   on the form factor evaluation of the $|V_{ub}|$ sensitive $\Lambda_b \to p l \nu$ decay  in the LCSR framework
\cite{Khodjamirian:2011jp} and from  lattice  with static $b$ quarks \cite{Detmold:2013nia}.

The  leptonic decay $ B \to \tau \nu$ can also provide information on $|V_{ub}|$. Previous  data have shown a disagreement of the measured branching ratio with the SM prediction, which  has softened significantly with the new data from Belle
Collaboration \cite{Adachi:2012mm}.

 In principle,  the method of extraction of $|V_{ub}|$  from inclusive $ \bar  B \rightarrow X_u  l \bar \nu_l$  decays  follows in the footsteps of the $|V_{cb}|$ determination from $ \bar  B \rightarrow X_c  l \bar \nu_l$, but the copious background from the
$ \bar B \rightarrow X_c l \bar \nu_l$  process, which has a rate about 50 times higher, limits
the experimental sensitivity to restricted regions of phase space,
where the  background  is  kinematically suppressed. The relative weight of  the threshold  region increases  and new
theoretical issues need to be addressed.
 Latest results by Belle \cite{Urquijo:2009tp} and BaBar   \cite{Lees:2011fv}
 access about the $ 90$\% of the $ \bar B \rightarrow X_u  l \bar \nu_l$ phase space.
On the theoretical side, several approaches have been devised
to analyze data in the threshold region,  with differences
in treatment of perturbative corrections and the
parameterization of nonperturbative effects.
\begin{table}[t]
\centering
\vskip 0.1 in
\begin{tabular}{l c} \hline \hline
{\it \bf Theory}  &   $|V_{ub}| \times 10^{3}$ \\
\hline
\hline
BLNP  & $ 4.40 \pm 0.15^{+0.19}_{-0.21}  $\\
\hline
DGE   & $4.45 \pm 0.15^{+ 0.15}_{- 0.16}$\\
\hline
ADFR   & $4.03 \pm 0.13^{+ 0.18}_{- 0.12}$\\
\hline
GGOU   & $4.39 \pm  0.15^{ + 0.12}_ { -0.20} $\\
\hline
\hline
\end{tabular}
\caption{ Comparison of  inclusive  determinations of $|V_{ub}|$ \cite{Amhis:2012bh}. }
\label{phidectab3}
\end{table}
%
The latest experimental analysis \cite{Lees:2011fv},  and  the HFAG averages \cite{Amhis:2012bh} rely on at least four different QCD calculations of the partial
decay rate:
 the
 BLNP approach
by Bosch, Lange, Neubert, and Paz \cite{Lange:2005yw}, the
GGOU one  by Gambino, Giordano, Ossola
and Uraltsev \cite{Gambino:2007rp},
the DGE one,
dressed gluon exponentiation, by Andersen and Gardi \cite{Andersen:2005mj, Gardi:2008bb} and  the ADFR approach, by Aglietti, Di Lodovico, Ferrara, and Ricciardi \cite{Aglietti:2004fz, Aglietti:2006yb, Aglietti:2007ik}.
These calculations take into account  the whole set of experimental results, or most of it, starting from 2002 CLEO data \cite{Bornheim:2002du}. Other theoretical approaches  have  been proposed in Refs.
\cite{Bauer:2001rc, Leibovich:1999xf, Ligeti:2008ac}.
The results  listed in  Table \ref{phidectab3}  give values in the range $\sim (3.9-4.6) \times 10^{-3}$, that
 are consistent within the errors, but  the theoretical uncertainty among determinations can reach 10\%.
In spite of all the experimental and theoretical efforts,
the values of $|V_{ub}|$ extracted from inclusive decays maintain about two $\sigma$ above the values given by exclusive determinations.
We can also compare with indirect fits,
$ |V_{ub}|  = (3.65 \pm  0.13) \times 10^{-3}$ by  UTfit \cite{Silvestrini} and
$ |V_{ub}|  = (3.49^{+0.21}_{-0.10}) \times 10^{-3}$ at $1 \sigma$ by CKMfitter \cite{CKMfitter}.
At variance
with the $|V_{cb}|$ case, the results of the global fit prefer a value for $|V_{ub}|$ that is closer to the
exclusive  determination.

\section{The $|V_{tb}|$ determination}
\label{vtb}

Ever since the existence of the  $b$-quark was inferred from the discovery of the $\Upsilon$ family of resonances at Fermilab in 1977 \cite{Herb:1977ek}, its weak isospin partner, the top quark, has been actively sought. For almost two decades, the top-quark eluded all direct searches at increasing higher energy, while indirect evidence of its existence and properties was provided by  loop mediated processes. Only  in 1995, eight years after the start of data taking at Tevatron, both CDF and \D0 Collaborations announced the observation of the top quark in $p \bar p \to t \bar t $ processes \cite{Abachi:1995iq,Abe:1995hr}.
At leading order in perturbation theory, there are two processes that contribute to $\bar t t $ production, quark-antiquark annihilation $\bar q q \to \bar t t$ and gluon-gluon fusion $\bar g g  \to \bar t t$. At the Tevatron the $\bar t t $
cross section is dominated by quark-antiquark annihilation, while at  LHC at $\sqrt{14}$ TeV, the situation is reversed.
The reason is that,
because of  the higher center of mass energy, at the LHC it is possible to produce $\bar t t $ already at lower $x$
of the incoming partons, where the gluon parton density dominates
over the quark densities. In addition, the Tevatron features antiquarks as constituent
quarks of the antiproton, leading to a considerable large antiquark density at large $x$.

As for the other CKM matrix elements,
the values of the top couplings   are not predicted within the SM, but $|V_{tb}|$  is expected to be close to unity as a consequence of unitarity and of the measured values for the other CKM elements. A recent global fit result gives   \cite{Silvestrini}   \beq  |V_{tb}| = 0.999106 \pm 0.000024 \label{SMvtb} \eeq
The large value of $|V_{tb}|$ in the SM   implies  that the ratio of branching fractions
\beq
R =\frac{B (t \to W b)}{ B(t \to W b)}= \frac{|V_{tb}|^2}{\sum_q |V_{tq}|^2} =   |V_{tb}|^2 \qquad (q=d, s,b)
\label{RSM}
\eeq
is close to unity, and that the top quark is expected to decay to a $W$ boson and a $b$ quark nearly 100\%  of the time.
In several models beyond the SM, different  assumptions, such as four quark generations and no CKM unitarity, hold, making a case for a measurement of $|V_{tb}|$ in the most direct possible way.

At hadron machines,  the top quark is mainly produced in $ t \bar t$ pairs via strong
interactions.  Around $56 \times 10^5$ $t \bar t$ pairs have been collected at ATLAS and CMS detectors,  about 8 times the number of  $t \bar t$ pairs  collected at Tevatron.
The measurement of the ratio $R$ is based on the number of jet tagged as $b$-jets for each $t \bar{t}$ event.
 The latest  \D0
results give \cite{Abazov:2011zk} $ R= 0.90 \pm 0.04$ (stat.+syst.),   which agrees within approximately
2.5 standard deviations with the SM prediction of $R$ close to one.
A simultaneous measurement
of $R= 0.94 \pm 0.09$ and $\sigma_{\bar t t}$ has been recently performed by the CDF collaboration;
they estimate \cite{Aaltonen:2013doa}
\beq
|V_{tb}| = 0.97 \pm 0.05
\eeq
assuming the SM relation (\ref{RSM}). At LHC at 7 TeV, the CMS collaboration has measured, in the dilepton channel, the value  $R= 0.98 \pm 0.04$ \cite{CMS:2012nua},  which is consistent with the SM prediction.

In strong interactions,
top quark is produced in pair; in electroweak interactions,  a single top can
 be produced. At hadron colliders, there are three channels of single top production at quark level, with different
cross sections strongly dependent on
the center of mass energy and the parton distribution function  of the incoming
partons. Apart from the $W t $ channel, which is the production of a  (close to) on-shell $W$ boson
and a top quark, via the gluonic fusion, there are the
$t$- and $s$-channels.
The $s$-channel ($q_1 \bar{q_2} \to t \bar{b}$) involves production of an off-shell, time-like
$W$
boson, which decays into a top and a bottom
quark.
The $t$-channel  ($q_1 (\bar{q_1}) b \to q_2 (\bar{q_2}) t$) mode is the exchange of a space-like $W$
boson between a light quark, and a bottom quark inside the incident hadrons, resulting in a jet and a single top quark.
Each mode has rather distinct event kinematics,
and thus they are observable separately from each other.
Single
top production was observed at the Tevatron based on a combination
of $t$- and $s$--channel processes  in 2009 \cite{Abazov:2009ii,Aaltonen:2009jj}, 14 yars after the top quark discovery, while  ATLAS and CMS Collaborations,
thanks to the much larger cross sections and better signal-over-background
available at the LHC, observed  single top already
 in 2011 and measured its cross section, in the $t$-channel, the year after \cite{Aad:2012ux,Chatrchyan:2012ep}.
ATLAS and CMS Collaborations have also reported first evidence of the $W t$ channel at the $ 4 \sigma$ level  \cite{Aad:2012xca, Chatrchyan:2012zca}.
Because of the massive particles in the final state, this mode has a
negligible rate at the Tevatron,  but not at the LHC,
where more partonic energy is available. At the opposite,
the $s$-channel has a  larger rate at the Tevatron than at the LHC,
because it is driven by
initial state anti-quark parton densities. There  is not yet evidence of this channel at LHC, but an upper
limit  has been set on the  production cross-section \cite{ATLAS:2011aia}.
At CDF, the cross section in the $s$-channel has been measured  with two-third of the data set \cite{CDF-10973}, and analyses are in progress to include  the whole data set.

The single top quark production cross section is directly proportional to the square of $|V_{tb}|$, allowing a direct measurement  of  $|V_{tb}|$ without assuming unitarity of the CKM matrix or three fermion generations. The extraction of $|V_{tb}|$ employs the following relation
\beq
| f_L  V_{tb} | =\sqrt{ \frac{ \sigma^{\mathrm{meas}}}{\sigma^{\mathrm{SM}}}}
\eeq
where $ \sigma^{\mathrm{meas}}$ and $ \sigma^{\mathrm{SM}}$ are the measured and the SM cross section in a specific channel, respectively, while $f_L$ takes into account a a possible left-handed anomalous coupling, being $f_L=1$ in the SM.
This equation  assumes that $|V_{tb}|$ is much larger than $|V_{ts}|$ and  $|V_{td}|$,
since the other CKM matrix elements could contribute to the top decay if they were not very small.
The current single top cross section measurements \cite{Abazov:2011pt, Chatrchyan:2012zca, CDF-10973,  CMSnote, Aad:2012xca, ATLASnote},
 have uncertainties at
the level of 10\%, the more precise being the
 CMS 7  TeV measurement  in the $t$-channel \cite{Chatrchyan:2012ep}, with
uncertainty of $\sim 5$\%, which yields
\beq
|V_{tb}| = 1.02 \pm 0.05 \pm 0.02
\eeq
assuming the SM value $f_L=1$.
Due to the large LHC statistics, single top measurements are
(mostly) systematics limited.
Dedicated strategies need to  be developed to increase precision and usefully employ this process
in the NP search.

Since unitarity of the $3 \times 3$ CKM matrix  strongly constraints $|V_{tb}|$,  deviations from the SM prediction in Eq. (\ref{SMvtb}) are expected from  NP that violate unitarity.
Deviations of the CKM unitarity  affect several observables, e.g. they may  lead to flavour-changing neutral currents
couplings which
are known from experiment to be severely suppressed; therefore unitarity violations need to be accordingly small.
A simple way to violate unitarity is enlarging the fermion sector, by including
a fourth quark generation or vector-like quarks, see e.g. \cite{Botella:2012ju,  Buras:2009ka, Alok:2010zj, Lacker:2012ek, Aguilar-Saavedra:2013qpa}.
Vector-like fermions are fermions that transform as triplets under
the colour gauge group and whose left- and right-handed chiralities belong to the same representation of the SM symmetry group $SU(3)_c \otimes SU(2)_L \otimes U(1)_Y$.
 In the SM with four  generations,  there
are no tree level flavour-changing neutral currents due to an exact GIM mechanism, while contributions from vector-like quarks may be suppressed by inverse ratios of their heavy masses.
Apart from dedicated models, vector-like quarks generically appears in a large number of extensions of the
SM, from composite and Little Higgs models to  Randall-Sundrum scenarios or E6 GUTs.
They have been actively sought, and
mass limits have  been provided by Tevatron and LHC \cite{Okada:2012gy}.
Other possible NP models  involve
flavour changing neutral currents  in the top quark sector mediated by the
$t$-channel exchange
of a new massive $Z^\prime$
boson, see e.g. \cite{Buras:2012jb}.
Experimental limits have also been set by both LHC and Tevatron on masses of new  $W^\prime$
bosons or charged Higgs bosons, whose existence would especially affect the single top
$s$-channel mode. Searches are also ongoing for single  excited $b^\ast$  quark production and decay to $W t$
\cite{Aad:2013rna}.
Excited quarks appear in several models,
for example in some Randall-Sundrum models
or in composite
Higgs models, see e.g. \cite{Vignaroli:2012si}.

\section*{Acknowledgments}

It is a pleasure to thank the Organizers of  Beauty 2013 for the invitation to a very interesting meeting and for providing a nice and stimulating atmosphere.

\bibliographystyle{PoS}
\bibliography{VxbRef}

\providecommand{\href}[2]{#2}\begingroup\raggedright\begin{thebibliography}{10}

\bibitem{Ricciardi:2013cda}
G.~Ricciardi, {\it {Determination of the CKM matrix elements $|V_{xb}|$}},
  {\em Mod. Phys. Lett.} {\bf A28} (2013) 1330016,
  [\href{http://xxx.lanl.gov/abs/1305.2844}{{\tt 1305.2844}}].

\bibitem{Ricciardi:2012pf}
G.~Ricciardi, {\it {Brief review on semileptonic B decays}},  {\em
  Mod.Phys.Lett.} {\bf A27} (2012) 1230037,
  [\href{http://xxx.lanl.gov/abs/1209.1407}{{\tt 1209.1407}}].

\bibitem{deDivitiis:2008df}
G.~de~Divitiis, R.~Petronzio, and N.~Tantalo, {\it {Quenched lattice
  calculation of the vector channel $B \to D^* l \nu$ decay rate}},  {\em
  Nucl.Phys.} {\bf B807} (2009) 373--395,
  [\href{http://xxx.lanl.gov/abs/0807.2944}{{\tt 0807.2944}}].

\bibitem{Aubert:2007rs}
{\bf BaBar} Collaboration, B.~Aubert {\em et~al.}, {\it {Determination of the
  form-factors for the decay $B^0 \to D^{*-} \ell^{+} \nu_{l}$ and of the CKM
  matrix element $|V_{cb}|$}},  {\em Phys.Rev.} {\bf D77} (2008) 032002,
  [\href{http://xxx.lanl.gov/abs/0705.4008}{{\tt 0705.4008}}].

\bibitem{deDivitiis:2007ui}
G.~de~Divitiis, E.~Molinaro, R.~Petronzio, and N.~Tantalo, {\it {Quenched
  lattice calculation of the $B \to D l \nu$ decay rate}},  {\em Phys.Lett.}
  {\bf B655} (2007) 45--49, [\href{http://xxx.lanl.gov/abs/0707.0582}{{\tt
  0707.0582}}].

\bibitem{deDivitiis:2007uk}
G.~de~Divitiis, R.~Petronzio, and N.~Tantalo, {\it {Quenched lattice
  calculation of semileptonic heavy-light meson form factors}},  {\em JHEP}
  {\bf 0710} (2007) 062, [\href{http://xxx.lanl.gov/abs/0707.0587}{{\tt
  0707.0587}}].

\bibitem{Aubert:2009ac}
{\bf BaBar} Collaboration, B.~Aubert {\em et~al.}, {\it {Measurement of |V(cb)|
  and the Form-Factor Slope in $ \bar B \to D l \bar \nu$ Decays in Events
  Tagged by a Fully Reconstructed B Meson}},  {\em Phys.Rev.Lett.} {\bf 104}
  (2010) 011802, [\href{http://xxx.lanl.gov/abs/0904.4063}{{\tt 0904.4063}}].

\bibitem{Amhis:2012bh}
{\bf Heavy Flavor Averaging Group} Collaboration, Y.~Amhis {\em et~al.}, {\it
  {Averages of B-Hadron, C-Hadron, and tau-lepton properties as of early
  2012}},  \href{http://xxx.lanl.gov/abs/1207.1158}{{\tt 1207.1158}}.

\bibitem{Bailey:2010gb}
{\bf Fermilab Lattice/MILC} Collaboration, J.~A. Bailey {\em et~al.}, {\it {$B
  \to D^* l \nu$ at zero recoil: an update}},  {\em PoS} {\bf LATTICE2010}
  (2010) 311, [\href{http://xxx.lanl.gov/abs/1011.2166}{{\tt 1011.2166}}].

\bibitem{Okamoto:2004xg}
M.~Okamoto, C.~Aubin, C.~Bernard, C.~E. DeTar, M.~Di~Pierro, {\em et~al.}, {\it
  {Semileptonic $ D \to \pi/K$ and $B \to \pi/D$ decays in 2+1 flavor lattice
  QCD}},  {\em Nucl.Phys.Proc.Suppl.} {\bf 140} (2005) 461--463,
  [\href{http://xxx.lanl.gov/abs/hep-lat/0409116}{{\tt hep-lat/0409116}}].

\bibitem{Laiho:2005ue}
J.~Laiho and R.~S. Van~de Water, {\it {$B \to D^\ast l \nu$ and $B \to D l \nu$
  form factors in staggered chiral perturbation theory.}},  {\em Phys.Rev.}
  {\bf D73} (2006) 054501, [\href{http://xxx.lanl.gov/abs/hep-lat/0512007}{{\tt
  hep-lat/0512007}}].

\bibitem{JackLaihotalk}
J.~Laiho, {\it {Lattice form factors and $| V_{cb}|$}}, . talk given at 7th
  International Workshop on the CKM Unitarity Triangle (CKM 2012), Sept 28-Oct
  2, Cincinnati, US.

\bibitem{Qiu:2012xi}
S.-W. Qiu, C.~DeTar, D.~Du, A.~S. Kronfeld, J.~Laiho, {\em et~al.}, {\it
  {Semileptonic B to D decays at nonzero recoil with 2+1 flavors of improved
  staggered quarks. An update}},  {\em PoS} {\bf LATTICE2012} (2012) 119,
  [\href{http://xxx.lanl.gov/abs/1211.2247}{{\tt 1211.2247}}].

\bibitem{Gambino:2012rd}
P.~Gambino, T.~Mannel, and N.~Uraltsev, {\it {$B \to D^*$ Zero-Recoil
  Formfactor and the Heavy Quark Expansion in QCD: A Systematic Study}},  {\em
  JHEP} {\bf 1210} (2012) 169, [\href{http://xxx.lanl.gov/abs/1206.2296}{{\tt
  1206.2296}}].

\bibitem{Uraltsev:2003ye}
N.~Uraltsev, {\it {A 'BPS' expansion for $B$ and $D$ mesons}},  {\em
  Phys.Lett.} {\bf B585} (2004) 253--262,
  [\href{http://xxx.lanl.gov/abs/hep-ph/0312001}{{\tt hep-ph/0312001}}].

\bibitem{Beringer:1900zz}
{\bf Particle Data Group} Collaboration, J.~Beringer {\em et~al.}, {\it {Review
  of Particle Physics (RPP)}},  {\em Phys.Rev.} {\bf D86} (2012) 010001.

\bibitem{Atoui:2013sca}
M.~Atoui, {\it {Lattice computation of $B \to D^*,\;D^{**} l \nu$ form factors
  at finite heavy masses}},  \href{http://xxx.lanl.gov/abs/1305.0462}{{\tt
  1305.0462}}. {in Proceedings of Rencontres de Moriond: QCD and High Energy
  Interactions, Mar 9-16, 2013 La Thuile, Italy}.

\bibitem{Aglietti:2002ew}
U.~Aglietti and G.~Ricciardi, {\it {Approximate NNLO threshold resummation in
  heavy flavor decays}},  {\em Phys.Rev.} {\bf D66} (2002) 074003,
  [\href{http://xxx.lanl.gov/abs/hep-ph/0204125}{{\tt hep-ph/0204125}}].

\bibitem{Aglietti:2005mb}
U.~Aglietti, G.~Ricciardi, and G.~Ferrera, {\it {Threshold resummed spectra in
  $ B \to X_u l \nu$ decays in NLO (I)}},  {\em Phys.Rev.} {\bf D74} (2006)
  034004, [\href{http://xxx.lanl.gov/abs/hep-ph/0507285}{{\tt
  hep-ph/0507285}}].

\bibitem{Aglietti:2005bm}
U.~Aglietti, G.~Ricciardi, and G.~Ferrera, {\it {Threshold resummed spectra in
  $B \to X_u l \nu$ decays in NLO (II)}},  {\em Phys.Rev.} {\bf D74} (2006)
  034005, [\href{http://xxx.lanl.gov/abs/hep-ph/0509095}{{\tt
  hep-ph/0509095}}].

\bibitem{Aglietti:2005eq}
U.~Aglietti, G.~Ricciardi, and G.~Ferrera, {\it {Threshold resummed spectra in
  $ B \to X_u l \nu$ decays in NLO (III).}},  {\em Phys.Rev.} {\bf D74} (2006)
  034006, [\href{http://xxx.lanl.gov/abs/hep-ph/0509271}{{\tt
  hep-ph/0509271}}].

\bibitem{Aglietti:2007bp}
U.~Aglietti, L.~Di~Giustino, G.~Ferrera, A.~Renzaglia, G.~Ricciardi, {\em
  et~al.}, {\it {Threshold Resummation in $B \to X_c l \nu_l $ Decays}},  {\em
  Phys.Lett.} {\bf B653} (2007) 38--52,
  [\href{http://xxx.lanl.gov/abs/0707.2010}{{\tt 0707.2010}}].

\bibitem{DiGiustino:2011jn}
L.~Di~Giustino, G.~Ricciardi, and L.~Trentadue, {\it {Minimal prescription
  corrected spectra in heavy quark decays}},  {\em Phys.Rev.} {\bf D84} (2011)
  034017, [\href{http://xxx.lanl.gov/abs/1102.0331}{{\tt 1102.0331}}].

\bibitem{Bailey:2008wp}
J.~A. Bailey, C.~Bernard, C.~E. DeTar, M.~Di~Pierro, A.~El-Khadra, {\em
  et~al.}, {\it {The $B \to \pi \ell \nu$ semileptonic form factor from
  three-flavor lattice QCD: A Model-independent determination of $|V_{ub}|$}},
  {\em Phys.Rev.} {\bf D79} (2009) 054507,
  [\href{http://xxx.lanl.gov/abs/0811.3640}{{\tt 0811.3640}}].

\bibitem{Dalgic:2006dt}
E.~Dalgic, A.~Gray, M.~Wingate, C.~T. Davies, G.~P. Lepage, {\em et~al.}, {\it
  {B meson semileptonic form-factors from unquenched lattice QCD}},  {\em
  Phys.Rev.} {\bf D73} (2006) 074502,
  [\href{http://xxx.lanl.gov/abs/hep-lat/0601021}{{\tt hep-lat/0601021}}].

\bibitem{AlHaydari:2009zr}
{\bf QCDSF} Collaboration, A.~Al-Haydari {\em et~al.}, {\it {Semileptonic form
  factors $D \to \pi, K$ and $B \to \pi, K$ from a fine lattice}},  {\em
  Eur.Phys.J.} {\bf A43} (2010) 107--120,
  [\href{http://xxx.lanl.gov/abs/0903.1664}{{\tt 0903.1664}}].

\bibitem{Lees:2012vv}
{\bf BaBar} Collaboration, J.~Lees {\em et~al.}, {\it {Branching fraction and
  form-factor shape measurements of exclusive charmless semileptonic B decays,
  and determination of $|V_{ub}|$}},  {\em Phys.Rev.} {\bf D86} (2012) 092004,
  [\href{http://xxx.lanl.gov/abs/1208.1253}{{\tt 1208.1253}}].

\bibitem{Ha:2010rf}
{\bf Belle} Collaboration, H.~Ha {\em et~al.}, {\it {Measurement of the decay
  $B^0\to\pi^-\ell^+\nu$ and determination of $|V_{ub}|$}},  {\em Phys.Rev.}
  {\bf D83} (2011) 071101, [\href{http://xxx.lanl.gov/abs/1012.0090}{{\tt
  1012.0090}}].

\bibitem{Khodjamirian:2011ub}
A.~Khodjamirian, T.~Mannel, N.~Offen, and Y.-M. Wang, {\it {$B \to \pi \ell
  \nu_l$ Width and $|V_{ub}|$ from QCD Light-Cone Sum Rules}},  {\em Phys.Rev.}
  {\bf D83} (2011) 094031, [\href{http://xxx.lanl.gov/abs/1103.2655}{{\tt
  1103.2655}}].

\bibitem{Bharucha:2012wy}
A.~Bharucha, {\it {Two-loop Corrections to the B to pi Form Factor from QCD Sum
  Rules on the Light-Cone and |V(ub)|}},  {\em JHEP} {\bf 1205} (2012) 092,
  [\href{http://xxx.lanl.gov/abs/1203.1359}{{\tt 1203.1359}}].

\bibitem{Li:2012gr}
Z.-H. Li, N.~Zhu, X.-J. Fan, and T.~Huang, {\it {Form Factors $f^{B\to
  \pi}_+(0)$ and $f^{D\to \pi}_+(0)$ in $QCD$ and Determination of $|V_{ub}|$
  and $|V_{cd}|$}},  {\em JHEP} {\bf 1205} (2012) 160,
  [\href{http://xxx.lanl.gov/abs/1206.0091}{{\tt 1206.0091}}].

\bibitem{Ball:2004ye}
P.~Ball and R.~Zwicky, {\it {New results on $B \to \pi, K, \eta$ decay
  formfactors from light-cone sum rules}},  {\em Phys.Rev.} {\bf D71} (2005)
  014015, [\href{http://xxx.lanl.gov/abs/hep-ph/0406232}{{\tt
  hep-ph/0406232}}].

\bibitem{delAmoSanchez:2010af}
{\bf BaBar} Collaboration, P.~del Amo~Sanchez {\em et~al.}, {\it {Study of $B
  \to \pi \ell \nu$ and $B \to \rho \ell \nu$ Decays and Determination of
  $|V_{ub}|$}},  {\em Phys.Rev.} {\bf D83} (2011) 032007,
  [\href{http://xxx.lanl.gov/abs/1005.3288}{{\tt 1005.3288}}].

\bibitem{delAmoSanchez:2010zd}
{\bf BaBar} Collaboration, P.~del Amo~Sanchez {\em et~al.}, {\it {Measurement
  of the $B^0 \to \pi^\ell \ell^+ \nu$ and $B^+ \to \eta^{(')} \ell^+ \nu$
  Branching Fractions, the $B^0 \to \pi^- \ell^+ \nu$ and $B^+ \to \eta \ell^+
  \nu$ Form-Factor Shapes, and Determination of $|V_{ub}|$}},  {\em Phys.Rev.}
  {\bf D83} (2011) 052011, [\href{http://xxx.lanl.gov/abs/1010.0987}{{\tt
  1010.0987}}].

\bibitem{DiDonato:2011kr}
C.~Di~Donato, G.~Ricciardi, and I.~Bigi, {\it {$\eta - \eta'$ Mixing - From
  electromagnetic transitions to weak decays of charm and beauty hadrons}},
  {\em Phys.Rev.} {\bf D85} (2012) 013016,
  [\href{http://xxx.lanl.gov/abs/1105.3557}{{\tt 1105.3557}}].

\bibitem{Khodjamirian:2011jp}
A.~Khodjamirian, C.~Klein, T.~Mannel, and Y.-M. Wang, {\it {Form Factors and
  Strong Couplings of Heavy Baryons from QCD Light-Cone Sum Rules}},  {\em
  JHEP} {\bf 1109} (2011) 106, [\href{http://xxx.lanl.gov/abs/1108.2971}{{\tt
  1108.2971}}].

\bibitem{Detmold:2013nia}
W.~Detmold, C.~J.~D. Lin, S.~Meinel, and M.~Wingate, {\it {$\Lambda_b \to p l^-
  \bar{\nu}$ form factors from lattice QCD with static b quarks}},
  \href{http://xxx.lanl.gov/abs/1306.0446}{{\tt 1306.0446}}.

\bibitem{Adachi:2012mm}
{\bf Belle Collaboration} Collaboration, I.~Adachi {\em et~al.}, {\it
  {Measurement of $B^- \to \tau^- \bar{\nu}_\tau$ with a Hadronic Tagging
  Method Using the Full Data Sample of Belle}},  {\em Phys.Rev.Lett.} {\bf 110}
  (2013) 131801, [\href{http://xxx.lanl.gov/abs/1208.4678}{{\tt 1208.4678}}].

\bibitem{Urquijo:2009tp}
{\bf Belle} Collaboration, P.~Urquijo {\em et~al.}, {\it {Measurement of
  $|V_{ub}|$ From Inclusive Charmless Semileptonic B Decays}},  {\em
  Phys.Rev.Lett.} {\bf 104} (2010) 021801,
  [\href{http://xxx.lanl.gov/abs/0907.0379}{{\tt 0907.0379}}].

\bibitem{Lees:2011fv}
{\bf BaBar} Collaboration, J.~Lees {\em et~al.}, {\it {Study of $\bar{B}\to X_u
  \ell \bar{\nu}$ decays in $B\bar{B}$ events tagged by a fully reconstructed
  B-meson decay and determination of $|V_{ub}|$}},  {\em Phys.Rev.} {\bf D86}
  (2012) 032004, [\href{http://xxx.lanl.gov/abs/1112.0702}{{\tt 1112.0702}}].

\bibitem{Lange:2005yw}
B.~O. Lange, M.~Neubert, and G.~Paz, {\it {Theory of charmless inclusive B
  decays and the extraction of V(ub)}},  {\em Phys.Rev.} {\bf D72} (2005)
  073006, [\href{http://xxx.lanl.gov/abs/hep-ph/0504071}{{\tt
  hep-ph/0504071}}].

\bibitem{Gambino:2007rp}
P.~Gambino, P.~Giordano, G.~Ossola, and N.~Uraltsev, {\it {Inclusive
  semileptonic B decays and the determination of |V(ub)|}},  {\em JHEP} {\bf
  0710} (2007) 058, [\href{http://xxx.lanl.gov/abs/0707.2493}{{\tt
  0707.2493}}].

\bibitem{Andersen:2005mj}
J.~R. Andersen and E.~Gardi, {\it {Inclusive spectra in charmless semileptonic
  B decays by dressed gluon exponentiation}},  {\em JHEP} {\bf 0601} (2006)
  097, [\href{http://xxx.lanl.gov/abs/hep-ph/0509360}{{\tt hep-ph/0509360}}].

\bibitem{Gardi:2008bb}
E.~Gardi, {\it {On the determination of |V(ub)| from inclusive semileptonic B
  decays}},  \href{http://xxx.lanl.gov/abs/0806.4524}{{\tt 0806.4524}}.

\bibitem{Aglietti:2004fz}
U.~Aglietti and G.~Ricciardi, {\it {A Model for next-to-leading order threshold
  resummed form-factors}},  {\em Phys.Rev.} {\bf D70} (2004) 114008,
  [\href{http://xxx.lanl.gov/abs/hep-ph/0407225}{{\tt hep-ph/0407225}}].

\bibitem{Aglietti:2006yb}
U.~Aglietti, G.~Ferrera, and G.~Ricciardi, {\it {Semi-Inclusive B Decays and a
  Model for Soft-Gluon Effects}},  {\em Nucl.Phys.} {\bf B768} (2007) 85--115,
  [\href{http://xxx.lanl.gov/abs/hep-ph/0608047}{{\tt hep-ph/0608047}}].

\bibitem{Aglietti:2007ik}
U.~Aglietti, F.~Di~Lodovico, G.~Ferrera, and G.~Ricciardi, {\it {Inclusive
  measure of |V(ub)| with the analytic coupling model}},  {\em Eur.Phys.J.}
  {\bf C59} (2009) 831--840, [\href{http://xxx.lanl.gov/abs/0711.0860}{{\tt
  0711.0860}}].

\bibitem{Bornheim:2002du}
{\bf CLEO} Collaboration, A.~Bornheim {\em et~al.}, {\it {Improved measurement
  of |V(ub)| with inclusive semileptonic B decays}},  {\em Phys.Rev.Lett.} {\bf
  88} (2002) 231803, [\href{http://xxx.lanl.gov/abs/hep-ex/0202019}{{\tt
  hep-ex/0202019}}].

\bibitem{Bauer:2001rc}
C.~W. Bauer, Z.~Ligeti, and M.~E. Luke, {\it {Precision determination of
  |V(ub)| from inclusive decays}},  {\em Phys.Rev.} {\bf D64} (2001) 113004,
  [\href{http://xxx.lanl.gov/abs/hep-ph/0107074}{{\tt hep-ph/0107074}}].

\bibitem{Leibovich:1999xf}
A.~K. Leibovich, I.~Low, and I.~Rothstein, {\it {Extracting V(ub) without
  recourse to structure functions}},  {\em Phys.Rev.} {\bf D61} (2000) 053006,
  [\href{http://xxx.lanl.gov/abs/hep-ph/9909404}{{\tt hep-ph/9909404}}].

\bibitem{Ligeti:2008ac}
Z.~Ligeti, I.~W. Stewart, and F.~J. Tackmann, {\it {Treating the b quark
  distribution function with reliable uncertainties}},  {\em Phys.Rev.} {\bf
  D78} (2008) 114014, [\href{http://xxx.lanl.gov/abs/0807.1926}{{\tt
  0807.1926}}].

\bibitem{Silvestrini}
L.~Silvestrini {Talk given at 14th International Conference on $B$ Physics at
  Hadron Machines (Beauty 2013), Apr 8-12, 2013, Bologna, Italy}, 2013.

\bibitem{CKMfitter}
{\bf CKMfitter} Collaboration, J.~Charles {\em et~al.} {\em Eur. Phys. J.} {\bf
  C41} (2005) 1, [\href{http://xxx.lanl.gov/abs/hep-ph/0406184}{{\tt
  hep-ph/0406184}}]. {Global fit results as of ICHEP 2012, updated results and
  plots available at: http://ckmfitter.in2p3.fr}.

\bibitem{Herb:1977ek}
S.~Herb, D.~Hom, L.~Lederman, J.~Sens, H.~Snyder, {\em et~al.}, {\it
  {Observation of a Dimuon Resonance at 9.5-GeV in 400-GeV Proton-Nucleus
  Collisions}},  {\em Phys.Rev.Lett.} {\bf 39} (1977) 252.

\bibitem{Abachi:1995iq}
{\bf \D0} Collaboration, S.~Abachi {\em et~al.}, {\it {Observation of the top
  quark}},  {\em Phys.Rev.Lett.} {\bf 74} (1995) 2632,
  [\href{http://xxx.lanl.gov/abs/hep-ex/9503003}{{\tt hep-ex/9503003}}].

\bibitem{Abe:1995hr}
{\bf CDF} Collaboration, F.~Abe {\em et~al.}, {\it {Observation of top quark
  production in $\bar{p}p$ collisions}},  {\em Phys.Rev.Lett.} {\bf 74} (1995)
  2626, [\href{http://xxx.lanl.gov/abs/hep-ex/9503002}{{\tt hep-ex/9503002}}].

\bibitem{Abazov:2011zk}
{\bf \D0} Collaboration, V.~Abazov {\em et~al.}, {\it {Precision measurement of
  the ratio ${\rm B}(t \to Wb)/{\rm B}(t \to Wq)$ and Extraction of $V_{tb}$}},
   {\em Phys.Rev.Lett.} {\bf 107} (2011) 121802,
  [\href{http://xxx.lanl.gov/abs/1106.5436}{{\tt 1106.5436}}].

\bibitem{Aaltonen:2013doa}
{\bf CDF} Collaboration, T.~Aaltonen {\em et~al.}, {\it {Measurement of R =
  \boldmath${B (t \rightarrow Wb)/B(t \rightarrow Wq)} $ in Top--quark--pair
  Decays using Lepton+jets Events and the Full CDF Run II Data set}},
  \href{http://xxx.lanl.gov/abs/1303.6142}{{\tt 1303.6142}}.

\bibitem{CMS:2012nua}
{\bf CMS} Collaboration, {\it {First measurement of $B(t \to Wb)/B(t \to Wq$)
  in the dilepton channel in pp collisions at $\sqrt{s}=7$ TeV}}, .
  {CMS-PAS-TOP-11-029}.

\bibitem{Abazov:2009ii}
{\bf \D0} Collaboration, V.~Abazov {\em et~al.}, {\it {Observation of Single
  Top Quark Production}},  {\em Phys.Rev.Lett.} {\bf 103} (2009) 092001,
  [\href{http://xxx.lanl.gov/abs/0903.0850}{{\tt 0903.0850}}].

\bibitem{Aaltonen:2009jj}
{\bf CDF} Collaboration, T.~Aaltonen {\em et~al.}, {\it {First Observation of
  Electroweak Single Top Quark Production}},  {\em Phys.Rev.Lett.} {\bf 103}
  (2009) 092002, [\href{http://xxx.lanl.gov/abs/0903.0885}{{\tt 0903.0885}}].

\bibitem{Aad:2012ux}
{\bf ATLAS} Collaboration, G.~Aad {\em et~al.}, {\it {Measurement of the
  $t$-channel single top-quark production cross section in $pp$ collisions at
  $\sqrt{s}=7$ TeV with the ATLAS detector}},  {\em Phys.Lett.} {\bf B717}
  (2012) 330, [\href{http://xxx.lanl.gov/abs/1205.3130}{{\tt 1205.3130}}].

\bibitem{Chatrchyan:2012ep}
{\bf CMS} Collaboration, S.~Chatrchyan {\em et~al.}, {\it {Measurement of the
  single-top-quark $t$-channel cross section in $pp$ collisions at $\sqrt{s}=7$
  TeV}},  {\em JHEP} {\bf 1212} (2012) 035,
  [\href{http://xxx.lanl.gov/abs/1209.4533}{{\tt 1209.4533}}].

\bibitem{Aad:2012xca}
{\bf ATLAS} Collaboration, G.~Aad {\em et~al.}, {\it {Evidence for the
  associated production of a $W$ boson and a top quark in ATLAS at $\sqrt{s}=7$
  TeV}},  {\em Phys.Lett.} {\bf B716} (2012) 142,
  [\href{http://xxx.lanl.gov/abs/1205.5764}{{\tt 1205.5764}}].

\bibitem{Chatrchyan:2012zca}
{\bf CMS} Collaboration, S.~Chatrchyan {\em et~al.}, {\it {Evidence for
  associated production of a single top quark and W boson in pp collisions at
  $\sqrt{s}=7$ TeV}},  {\em Phys.Rev.Lett.} {\bf 110} (2013) 022003,
  [\href{http://xxx.lanl.gov/abs/1209.3489}{{\tt 1209.3489}}].

\bibitem{ATLAS:2011aia}
{\bf ATLAS} Collaboration, {\it {Search for s-Channel Single Top-Quark
  Production in $pp$ Collisions at $\sqrt{s}$ = 7 TeV}}, .
  {ATLAS-CONF-2011-118}.

\bibitem{CDF-10973}
{\bf CDF} Collaboration, CDF {public note CDF-10793}, 2012.

\bibitem{Abazov:2011pt}
{\bf \D0} Collaboration, V.~M. Abazov {\em et~al.}, {\it {Measurements of
  single top quark production cross sections and $|V_{tb}|$ in $p\bar{p}$
  collisions at $\sqrt{s}=1.96$ TeV}},  {\em Phys.Rev.} {\bf D84} (2011)
  112001, [\href{http://xxx.lanl.gov/abs/1108.3091}{{\tt 1108.3091}}].

\bibitem{CMSnote}
{\bf CMS} Collaboration, CMS { PAS TOP-12-011
  http://cds.cern.ch/record/1478935}, 2012.

\bibitem{ATLASnote}
{\bf ATLAS} Collaboration, ATLAS {CONF-2012-132
  http://cds.cern.ch/record/1478371}, 2012.

\bibitem{Botella:2012ju}
F.~Botella, G.~Branco, and M.~Nebot, {\it {The Hunt for New Physics in the
  Flavour Sector with up vector-like quarks}},  {\em JHEP} {\bf 1212} (2012)
  040, [\href{http://xxx.lanl.gov/abs/1207.4440}{{\tt 1207.4440}}].

\bibitem{Buras:2009ka}
A.~J. Buras, B.~Duling, and S.~Gori, {\it {The Impact of Kaluza-Klein Fermions
  on Standard Model Fermion Couplings in a RS Model with Custodial
  Protection}},  {\em JHEP} {\bf 0909} (2009) 076,
  [\href{http://xxx.lanl.gov/abs/0905.2318}{{\tt 0905.2318}}].

\bibitem{Alok:2010zj}
A.~K. Alok, A.~Dighe, and D.~London, {\it {Constraints on the Four-Generation
  Quark Mixing Matrix from a Fit to Flavor-Physics Data}},  {\em Phys.Rev.}
  {\bf D83} (2011) 073008, [\href{http://xxx.lanl.gov/abs/1011.2634}{{\tt
  1011.2634}}].

\bibitem{Lacker:2012ek}
H.~Lacker, A.~Menzel, F.~Spettel, D.~Hirschbuhl, J.~Luck, {\em et~al.}, {\it
  {Model-independent extraction of $|V_{tq}|$ matrix elements from top-quark
  measurements at hadron colliders}},  {\em Eur.Phys.J.} {\bf C72} (2012) 2048,
  [\href{http://xxx.lanl.gov/abs/1202.4694}{{\tt 1202.4694}}].

\bibitem{Aguilar-Saavedra:2013qpa}
J.~Aguilar-Saavedra, R.~Benbrik, S.~Heinemeyer, and M.~Perez-Victoria, {\it {A
  handbook of vector-like quarks: mixing and single production}},
  \href{http://xxx.lanl.gov/abs/1306.0572}{{\tt 1306.0572}}.

\bibitem{Okada:2012gy}
Y.~Okada and L.~Panizzi, {\it {LHC signatures of vector-like quarks}},  {\em
  Adv.High Energy Phys.} {\bf 2013} (2013) 364936,
  [\href{http://xxx.lanl.gov/abs/1207.5607}{{\tt 1207.5607}}].

\bibitem{Buras:2012jb}
A.~J. Buras, F.~De~Fazio, and J.~Girrbach, {\it {The Anatomy of Z' and Z with
  Flavour Changing Neutral Currents in the Flavour Precision Era}},  {\em JHEP}
  {\bf 1302} (2013) 116, [\href{http://xxx.lanl.gov/abs/1211.1896}{{\tt
  1211.1896}}].

\bibitem{Aad:2013rna}
{\bf ATLAS} Collaboration, G.~Aad {\em et~al.}, {\it {Search for single
  $b^*$-quark production with the ATLAS detector at $\sqrt{s}=7$ TeV}},  {\em
  Phys.Lett.} {\bf B721} (2013) 171--189,
  [\href{http://xxx.lanl.gov/abs/1301.1583}{{\tt 1301.1583}}].

\bibitem{Vignaroli:2012si}
N.~Vignaroli, {\it {$\Delta$ F=1 constraints on composite Higgs models with LR
  parity}},  {\em Phys.Rev.} {\bf D86} (2012) 115011,
  [\href{http://xxx.lanl.gov/abs/1204.0478}{{\tt 1204.0478}}].

\end{thebibliography}\endgroup

\end{document}